\documentstyle[12pt,aaspp4,psfig,flushrt]{article} 
\begin{document}

\title{A Search for Optical Pulsations from \\ Two Young Southern Pulsars}

\author{Deepto~Chakrabarty\altaffilmark{1,2} and Victoria~M.~Kaspi}
\affil{\footnotesize Center for Space Research and Department of Physics, 
  Massachusetts Institute of Technology, \\ Cambridge, MA 02139; 
  deepto@space.mit.edu, vicky@space.mit.edu}

\altaffiltext{1}{NASA Compton GRO Postdoctoral Fellow}
\altaffiltext{2}{Visiting Astronomer, Cerro Tololo Inter-American
Observatory, National Optical Astronomy Observatories, which are
operated by the Associated Universities for Research in Astronomy
under contract with the National Science Foundation.}

\medskip
\centerline{\footnotesize\em Submitted on 1998 January 5; accepted 1998 March 5}
\centerline{To appear in \sc The Astrophysical Journal Letters}

\begin{abstract}

We report on high-speed optical photometry of the radio positions of
two young rotation-powered pulsars.  No pulsations were detected from
the optical counterpart proposed by Caraveo et al. (1994) for PSR
B1509--58, with a 2$\sigma$ upper limit on the pulsed fraction of $<
12$\%, significantly lower than that measured in the five known
optical pulsars.  Given its low pulsed fraction, high optical 
luminosity, and significant (8\%) chance coincidence probability, we
suggest that this candidate is not associated with the pulsar.  We
also find that the still-unidentified optical counterpart of PSR
B1706--44 has $R\gtrsim 18$ and lies within 3 arcsec of an $R=16.6$ star.

\end{abstract}


\section{INTRODUCTION} 

Nearly 800 isolated rotation-powered neutron stars are currently
known, most of them discovered through their radio pulsations (Taylor,
Manchester, \& Lyne 1993; Taylor et al. 1995).  A small fraction of
these pulsars has also been detected in the optical, X-ray, and
gamma-ray bands; in some cases, this emission is also pulsed.  The
radiation from these objects is believed to arise from three distinct
processes.  In the youngest pulsars (as estimated by their spindown
age, $\tau\equiv P/2\dot P\lesssim 10^4$ yr, where $P$ and $\dot P$
are the pulsar's spin period and period derivative), non-thermal
emission arising from the relativistic acceleration of $e^+e^-$ pairs
in the corotating magnetosphere dominates at all wavelengths (see
Romani 1996).  In young and middle-aged pulsars ($10^4 \lesssim \tau
\lesssim 10^6$ yr), thermal soft X-ray emission arising from the
initial cooling of the neutron star surface has been detected, in
addition to non-thermal magnetospheric emission (see \"Ogelman 1993).
In two nearby older pulsars ($10^6 \lesssim \tau \lesssim
10^7$ yr), thermal soft X-ray emission is detected and is thought to
arise from heating of the polar caps by returning $e^+e^-$
pairs accelerated in the outer magnetosphere (Yancopoulos, Hamilton,
\& Helfand 1994; Manning \& Wilmore 1994).  The other known
rotation-powered pulsars are thus far detected only through their
non-thermal radio emission.  Optical/ultraviolet emission of young and
middle-aged pulsars probes both the non-thermal and thermal
components and thus may provide significant constraints on theoretical
models when combined with X-ray data.

Optical emission has been reported from directions of nine pulsars.
Pulsations were detected from all five pulsars for which time-resolved
data are available, most recently from PSR B0656+14 and Geminga
(Shearer et al. 1997, 1998).  In most cases, the optical pulsed
fraction is $\sim$100\%.  Because the intrinsically faint optical
emission expected from pulsars leads to non-negligible probabilities
of chance coincidence with the radio position, the detection of
optical pulsations is a crucial test of a proposed counterpart.

In this paper, we report negative results of a search for optical
pulsations near the radio positions of two young pulsars in the
Southern sky.  These non-detections call into question the counterpart
proposed by Caraveo, Mereghetti, \& Bignami (1994, hereafter CMB94)
for the very young PSR B1509--58 ($P=150$ ms, $\dot P = 1.6\times
10^{-12}$, $\tau=1600$ yr) and set the first limits on an optical
counterpart for the relatively young PSR B1706--44 ($P=102$ ms, $\dot
P=9.3\times 10^{-14}$, $\tau=17400$ yr).  In the course of completing
this work, we learned of two independent non-detections of pulsations
from the proposed optical counterpart of PSR B1509--58 (R.~N. Manchester
1997, private communication; Mignani et al. 1998).

\section{OBSERVATIONS AND ANALYSIS}

We obtained high-speed photoelectric photometry of the radio coordinates of
the two young pulsars on UT 1995 May 26--29 (MJD 49863--49866) with the
4-m Blanco telescope at the Cerro Tololo Inter-American Observatory
(CTIO) in La Serena, Chile.  The details of the observational setup
and the photometric and timing calibrations are described by
Chakrabarty (1998).  The night of May 26 was photometric, but sky
conditions gradually deteriorated over the next several nights, with
many high, thin clouds present on May 29. 

Our observations were made with the Automated Single Channel Aperture
Photometer (ASCAP) and a dry-ice-cooled Varian VPM-159A
photomultiplier tube at the $f/7.5$ Ritchey-Chretien focus. The Varian
phototube, which has an InGaAsP photocathode, is sensitive over a wide
wavelength range, with a quantum efficiency of 15\% at 4000~\AA\ and
5\% at 9000~\AA.  Most of the observations used the $R$ filter from
the {\em UBVRI} filter set described by Graham (1982).  Two of the
observations used a longpass RG-610 filter, which blocks optical
wavelengths shortward of 6100~\AA.  Most of the observations were made
through a 6.6~arcsec circular aperture, although one observation (run
45R) was made through a 9.2~arcsec aperture.  The data were recorded
at 1 ms resolution and rebinned to 10 ms resolution for timing
analysis.  One of the observations (run 45R) contained an instrumental
signal at 60 Hz and its higher harmonics along with another at 1.75 Hz
and its higher harmonics.  These contaminants were restricted to a
discrete set of frequencies and were easily identified.  Table~1 gives
a log of our observations.

We used the radio coordinates of the two pulsars to position
our aperture (see Table 2).  For PSR B1509--58, we observed the
optical counterpart proposed by CMB94, which is coincident with the
radio timing position of Kaspi et al. (1994) within 1
arcsec\footnote{CMB94 used the radio timing position of Manchester,
Durdin, \& Newton (1985), as quoted in the catalog of Taylor et
al. (1993).  The Kaspi et al. (1994) position supersedes this but has
a somewhat larger uncertainty owing to the inclusion of systematic
errors due to timing noise.}.  For PSR B1706--44, there are two new
positions available due to radio timing (Wang et al. 1998) and radio
interferometry (Frail \& Goss 1998) which are essentially identical.
This position is 2.7~arcsec from a relatively bright optical star,
which we designate as star~1.  This and several other stars, along
with the radio position, are shown in Figure~1, which is a 2-min
$R$-band exposure of the field taken from the CTIO 1.5-m telescope on
1995 May 28.  We computed an astrometric solution for this image using
stellar  coordinates derived from a 1987 $V$-band plate of the region, obtained
from the STScI Digitized Sky Survey\footnote{Based on photographic
data obtained using the UK Schmidt Telescope, operated by the Royal
Observatory, Edinburgh with funding from the UK SERC.  The Digitized
Sky Survey was produced at STScI under NASA grant NAGW-2166.}.  We
observed three different positions in this field, indicated by the
circles on Figure~1.  The first two positions were centered on stars~1
and 2, respectively, and used a 6.6~arcsec aperture.  The third
position was centered near stars~4 and 5, and used a 9.2~arcsec
aperture.  Only the aperture centered on star~1 included the new radio
position. 

\section{RESULTS}

For each observation, the data were transformed to the solar
system barycenter frame using the Jet Propulsion Laboratory DE200
solar system ephemeris (Standish et al. 1992), and the time series was
flattened by subtracting a low-order polynomial.  The pulse phases for
the binned arrival times were computed according to a contemporaneous
radio timing model of the form 
\begin{equation}
\phi(t) = \nu_0(t-t_0) + \frac{1}{2}\dot\nu_0(t-t_0)^2
   + \frac{1}{6}\ddot\nu_0(t-t_0)^3 ,
\end{equation}
where $\phi$ is the pulse phase as a function of the barycentric time
$t$; $t_0$ is the epoch for the timing model; and $\nu_0$,
$\dot\nu_0$, and $\ddot\nu_0$ are the pulse frequency and its
derivatives at $t=t_0$.  The radio timing parameters for PSRs B1509--58
and B1706--44 are regularly monitored at Parkes Observatory in
Australia and are available in an electronic database maintained at
Princeton University\footnote{Available on the web at
http://pulsar.princeton.edu/ftp/gro.  The parameters for these two
pulsars are based on observations at the Parkes radio telescope by
R.~N. Manchester, M. Bailes, and collaborators.}.

The resulting pulse phase distributions were searched for evidence of
periodicity using the $H$-test, a sensitive statistical test for weak
periodic signals of unknown pulse shape (de Jager, Swanepoel, \&
Raubenheimer 1989).  The chief advantages of the $H$-test are that it
is independent of binning and that it is sensitive to a wide range of
pulse shapes.   We found no convincing evidence for pulsations in any
of our observations.  The computed probabilities for a uniform
(unpulsed) distribution of phases are included in Table 1.  We also
used the background-subtracted unpulsed (``DC'') intensity 
to estimate a 95\%-confidence upper limit on the pulsed
fraction (see de Jager 1994, Chakrabarty et al. 1995, and Chakrabarty
1996 for a discussion of this problem).   These upper limits are
included in Table~1. 

\section{DISCUSSION}

Based on its radio dispersion measure and the distance model of Taylor
\& Cordes (1993), PSR B1509--58 is estimated to be 4.4 kpc distant.
Assuming a rough $V$ extinction law of $\sim 1$ mag kpc$^{-1}$ (e.g.,
Spitzer 1978), this implies an absolute magnitude of $M_V=4.4$ for the
CMB94 candidate counterpart.  As CMB94 pointed out, this optical
luminosity far exceeds the prediction of the $B^4P^{-10}$ law derived
by assuming synchrotron emission near the light cylinder and scaling
the surface dipole magnetic field and spin period of the Crab pulsar
(Pacini 1971).  It also far exceeds an extrapolation of the pulsar's
non-thermal power-law X-ray spectrum into the optical.  The Pacini
scaling law is able to account for the optical luminosity of the other
young pulsars where non-thermal emission dominates (Pacini \& Salvati
1987), but it cannot account for the older pulsars, whose optical
emission is probably thermal in origin (e.g., Shearer et al. 1997).
However, since non-thermal emission should dominate in PSR B1509--58,
one would expect the scaling law to apply. 

Our 2$\sigma$ upper limit of $<12$\% on the optical pulsed fraction of
the CMB94 candidate contrasts with the $\sim100$\% optical pulsed fraction
observed in 4 of the 5 known optical pulsars (see Table 3).  Only PSR
B0540--69 has a lower pulsed fraction of 16\% (Boyd et al. 1995).
However, this value must be regarded as a lower limit due to the
uncertain contribution from the synchrotron nebula enshrouding the
pulsar.  At the distance of the LMC, the 0.65 arcsec aperture used in
the {\em Hubble Space Telescope (HST)} observation covered a 0.16 pc
region around PSR B0540--69 (Boyd et al. 1995).  By comparison, the 1
arcsec aperture used in an {\em HST} observation of the much closer
Crab pulsar covered only 0.01 pc around the neutron star and still
included a $\sim5$\% nebular contribution (Percival et al. 1993).
Therefore, the optical pulsed fraction of PSR B0540--69 is probably
considerably higher than 16\%.  Thus the CMB94 candidate, if it were
the true pulsar counterpart, would be remarkable both for its high
optical efficiency and its low optical pulsed fraction.  

It is instructive to compute the probability of a chance optical
coincidence with the radio position of PSR B1509--58.  The density of
stars with $V<22$ in the Galactic plane is $\sim 10^{5.5}$
stars~deg$^{-2}$ (Allen 1973).  Poisson statistics then give a
probability of $\sim8$\% for finding an unrelated $V\leq22$ star
within 1 arcsec of the pulsar.  Given its high optical luminosity,
small pulsed fraction, and significant chance coincidence probability,
we suggest that the CMB94 candidate is not associated with
PSR B1509--58.  Deep optical spectroscopy should eventually settle the
question.  If it is not the true counterpart, then identification of
the actual $V\gtrsim22$ counterpart at such a necessarily small angular
separation from the CMB94 candidate is likely impossible with
ground-based observations.  

We can use our non-detection of pulsations from the observed stars in
the PSR B1706--44 field to place the first limits on this pulsar's
optical flux.  Our photoelectric photometry of star~1 on May 26
yielded the following magnitude and colors: $V=17.3$, $U-B=1.0$,
$B-V=1.1$, $V-R=0.7$, and $R-I=0.4$.  Based on differential photometry
using our $R$-band image of the field, we conclude that $R\gtrsim18$.
Given the nearby presence of star~1, ground-based
identification of this optical counterpart will also be difficult.
The present optical limits are not constraining for the Pacini (1971)
relation, which predicts a dereddened magnitude of $R\sim 28$ for PSR
B1706--44.

\acknowledgements{We thank Ahren Lembke-Windler for background
  research for this paper, Paul Eskridge for acquiring a CCD image of
  the PSR B1706--44 field for us, Dick Manchester for a careful
  reading of the manuscript, and Dale Frail and Roberto Mignani for discussing
  their work with us prior to publication.  D.C. thanks Andy
  Layden, Ramon Galvez, Patricio Ugarte, and the CTIO staff for their
  assistance and excellent support, and acknowledges a thesis
  observing travel award from CTIO.  D.C. was also supported by a NASA
  GSRP Graduate Fellowship at Caltech under grant NGT-51184, and by a
  NASA Compton GRO Postdoctoral Fellowship at MIT under grant NAG
  5-3109.  V.M.K. was supported in part by a Hubble Postdoctoral
  Fellowship under grant HF-1061.01-94A from STScI, which is operated
  by AURA under NASA contract NAS 5-26555.}

\pagebreak

{\small
\begin{deluxetable}{llcclcl}
\tablewidth{6.8in}
\tablecaption{Log of Observations}
\tablecolumns{7}
\tablehead{
 & &  & \colhead{Duration} & & \colhead{Chance} & \colhead{Pulsed}\\
\colhead{Target} & \colhead{Run ID} & \colhead{MJD\tablenotemark{a}} & 
  \colhead{(min)}& \colhead{Filter} & 
  \colhead{probability\tablenotemark{b}} & 
  \colhead{fraction\tablenotemark{c}} }
\startdata
\sidehead{\em PSR B1509--58}
CMB94 candidate & 33R & 49865.145 & 49 & $R$ & 40.1\% & $<12$\% \\
CMB94 candidate & 41R & 49866.010 & 60 & $R$ & 4.6\% & $<17$\% \\
\sidehead{\em PSR B1706--44}
Star 1 & 26RG & 49864.306 & 60 & RG-610 & 38.9\% & $<1$\% \\
Star 1 & 34RG & 49865.207 & 108 & RG-610 & 10.2\% & $<0.2$\% \\
Star 2 & 44R & 49866.124 & 30 & $R$ & 29.1\% & $<3$\% \\
Stars 4 and 5 & 45R & 49866.152 & 120 & $R$ & 9.0\% & $<6$\% \\
\enddata
\tablenotetext{a}{Modified Julian date = JD -- 2,400,000.5.  MJD 49864
= 1995 May 27.}
\tablenotetext{b}{$H$-test probability of a uniform distribution.}
\tablenotetext{c}{95\%-confidence upper limits.}
\end{deluxetable}
}

{\small
\begin{deluxetable}{lllc}
\tablewidth{4.5in}
\tablecaption{J2000.0 Positions}
\tablecolumns{4}
\tablehead{
\colhead{Object} &\colhead{RA (h m s)} & \colhead{Decl. 
   ($^\circ$\ \arcmin\ \arcsec)}&
   \colhead{Ref.}}
\startdata
\sidehead{\em PSR B1509--58}
Radio timing position     & 15 13 55.62(9) & --59 08 09(1) & 1 \\
CMB94 optical candidate\tablenotemark{a} & 15 13 55.52 & --59 08 08.8 & 2 \\
\sidehead{\em PSR B1706--44}
Radio interferometry     & 17 09 42.732(24) & --44 29 07.7(8) & 3 \\
Radio timing position    & 17 09 42.73(2) & --44 29 07.7(6) & 4 \\
Star 1                   & 17 09 42.80(6) & --44 29 10.3(7) & 5 \\
\enddata
\tablenotetext{a}{0.5 arcsec (1$\sigma$) error radius.}
\tablerefs{(1) Kaspi et al. 1994; (2) CMB94; (3) Frail \& Goss 1998; 
(4) Wang et al. 1998; (5) This work.}
\end{deluxetable}
}

{\small
\begin{deluxetable}{lcrrl}
\tablewidth{5in}
\tablecaption{Optical and X-Ray Pulsed Fractions}
\tablecolumns{5}
\tablehead{
 &\colhead{log $\tau$} &\multicolumn{2}{c}{Pulsed fraction\tablenotemark{a}}&\\
\cline{3-4}
\colhead{Pulsar} & \colhead{(yr)} & \colhead{Optical} & 
   \colhead{X-ray\tablenotemark{b}} & \colhead{References}}
\startdata
\sidehead{\em Very young pulsars}
Crab pulsar & 3.1 & 100\%  & $\gtrsim 75$\%  & 1, 2 \\
PSR B0540--69 & 3.2 &  $>16$\%  &          15\%   & 2, 3\\
PSR B1509--58 & 3.2 & \nodata & 65\%   & 2 \\
\sidehead{\em Young pulsars}
Vela pulsar & 4.1 & $\sim100$\%  &  11\%     & 2, 4\\
PSR B1706--44 & 4.2 & \nodata & $\lesssim 18$\% & 5\\
\sidehead{\em Middle-aged pulsars}
PSR B0656+14  & 5.0 & $\sim100$\%  & 18\% & 2, 6 \\
Geminga   & 5.5 & $\gtrsim70$\% & 33\%  & 2, 7\\
PSR B1055--52 & 5.7 & ?           & 17\%  & 2, 8\\
\sidehead{\em Old pulsars}
PSR B1929+10  & 6.5 & ?           & 30\% & 2, 9\\
PSR B0950+08  & 7.2 & ?           & ?    & 2, 9\\
\enddata
\tablenotetext{a}{A question mark indicates an integrated detection
with no pulsation search.  A blank entry indicates lack of a counterpart.}
\tablenotetext{b}{As measured in the 0.1--2.4 keV band by {\em ROSAT}.}
\tablerefs{(1) Percival et al. 1993; (2) Becker \& Tr\"umper 1997 and
references therein; (3) Boyd et al. 1995; (4) Wallace
et al. 1977, Manchester et al. 1978, Nasuti et al. 1997; (5) Becker et
al. 1995; (6) Shearer et al. 1997; (7) Shearer et al. 1998;
(8) Mignani et al. 1997; (9) Pavlov et al. 1996.
}
\end{deluxetable}
}

\clearpage
\centerline{FIGURE CAPTION}

\noindent 
FIGURE 1: $R$-band image of the PSR B1706--44 field.  North is up, and
east is to the left. The cross, which shows the radio position for the
pulsar (Wang et al. 1998; Frail \& Goss 1998), is drawn 2 arcsec
across.  This is larger than the $\lesssim$ 1~arcsec measurement
uncertainty.  The three circles indicate the aperture position
used in our observations.  Only the central aperture contained the
new radio position.  Star 1 has $R=16.6$ and star 5 has
$R\approx 20.9$.  The brightest star in the field is HD 329564, with
$R\lesssim 13.4$.

\pagebreak
\pagestyle{empty}
\thispagestyle{empty}
\begin{figure}
\centerline{Figure 1}
\centerline{\psfig{file=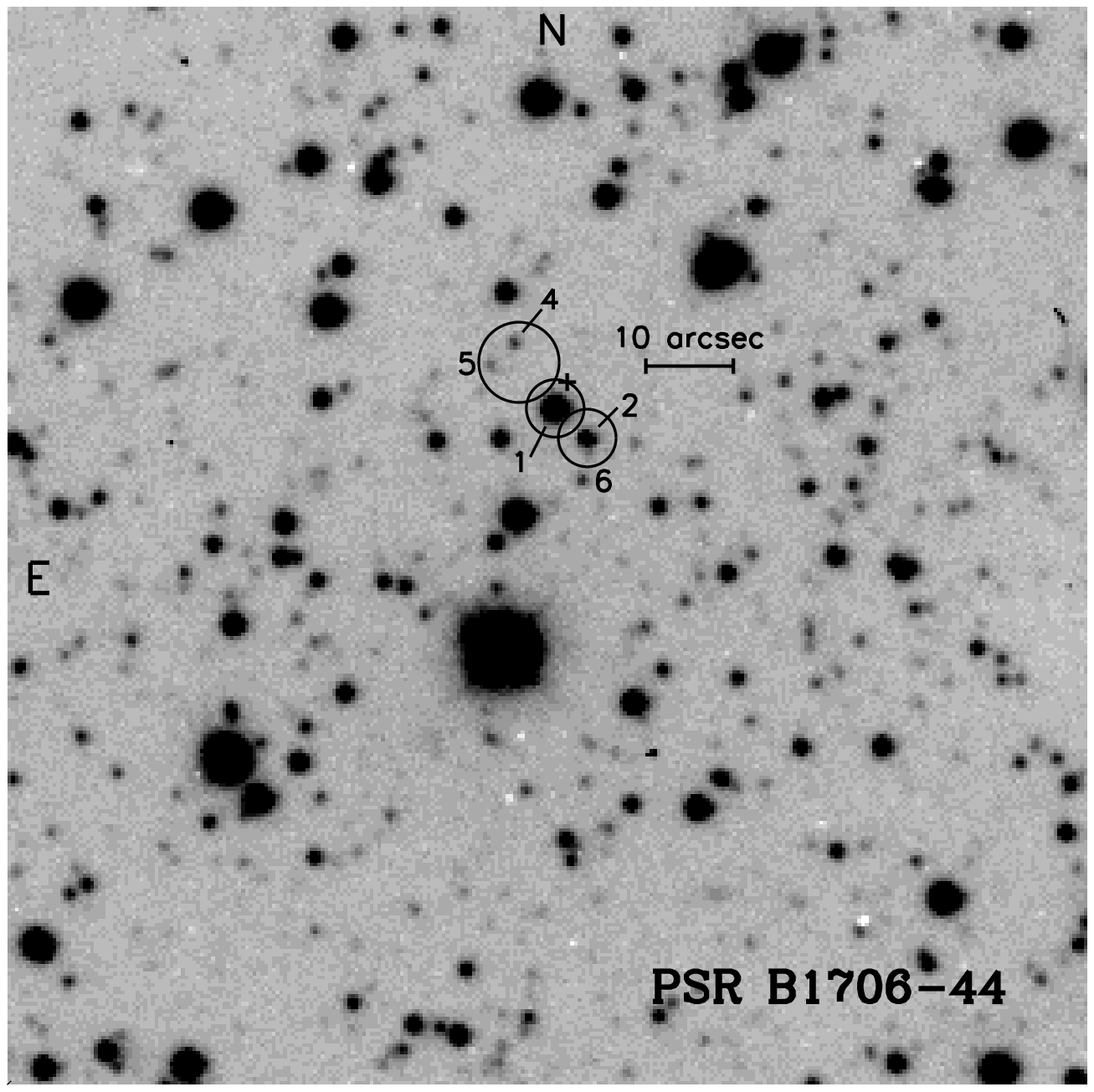}}
\end{figure}

\end{document}